\documentclass[12pt]{article}
\input epsf.sty
\usepackage{mathrsfs}
\usepackage{amsmath}
\usepackage{mathrsfs}
\usepackage{amssymb}
\usepackage{color}
\usepackage{psfrag}
\usepackage{graphicx}
\usepackage{subfigure}
\usepackage{rotating}

\newcommand{\bea}{\begin{eqnarray}}
\newcommand{\eea}{\end{eqnarray}}
\newcommand{\be}{\begin{equation}}
\newcommand{\ee}{\end{equation}}
\newcommand{\vs}[1]{\vspace{#1 mm}}

\newcommand{\dsl}{\pa \kern-0.5em /}

\newcommand{\half}{\frac{1}{2}}
\newcommand{\pa}{\partial}

\newcommand{\nn}{\nonumber\\}

\begin{document}
\topmargin 0pt
\oddsidemargin 0mm

\begin{flushright}



\end{flushright}

\vspace{2mm}

\begin{center}
{\Large \bf Lifshitz-like space-time from intersecting branes 
\\in string/M theory}

\vs{10}

{Parijat Dey\footnote{E-mail: parijat.dey@saha.ac.in} and 
Shibaji Roy\footnote{E-mail: shibaji.roy@saha.ac.in}}

 \vspace{4mm}

{\em

 Saha Institute of Nuclear Physics,
 1/AF Bidhannagar, Calcutta-700 064, India\\}

\end{center}

\vs{10}

\begin{abstract}
We construct 1/4 BPS, threshold F-D$p$ bound states (with $0\leq p \leq 5$) of 
type II string theories by applying S- and T-dualities to the D1-D5 system of
type IIB string theory. These are different from the known 1/2 BPS, 
non-threshold F-D$p$ bound states. The near horizon limits of these solutions 
yield Lifshitz-like space-times
with varying dynamical critical exponent $z=2(5-p)/(4-p)$, for $p\neq 4$, 
along with the hyperscaling violation exponent $\theta = p - (p-2)/(4-p)$, 
showing how Lifshitz-like space-time can be obtained from string 
theory. The dilatons are in general non-constant (except 
for $p=1$). We discuss the holographic RG flows and the phase structures
of these solutions. For $p=4$, we do not get a Lifshitz-like 
space-time, but the near horizon limit in this case leads to an AdS$_2$ 
space. 
\end{abstract}

\newpage

\section{Introduction}

Holographic ideas \cite{'tHooft:1993gx} in the form of gauge/gravity 
duality \cite{Maldacena:1997re} have been proved quite
useful in recent years to understand the strong coupling behavior of theories
without gravity from the weakly coupled gravity theories in one higher
space-time dimensions. This general idea is believed to be applicable not only 
to relativistic theories suitable for QCD (see \cite{Erlich:2008en}, for 
some reviews), but also to non-relativistic theories suitable for condensed 
matter systems (for reviews, see \cite{Hartnoll:2009sz}). 

Non-relativistic symmetries
can be of two types, namely, the Schr\" odinger symmetry and the Lifshitz
symmetry.  In both types the time and spaces scale differently breaking the 
Lorentz invariance. Schr\" odinger symmetry consists of time and space
translations, spatial rotations, Galilean boosts, dilatations or scaling
symmetry, a special conformal transformation and a particle number symmetry.  
On the other hand, Lifshitz symmetry is a much smaller symmetry with only
time and space translations, spatial rotations and a scaling symmetry.  
Gravitational theories having Sch\" odinger symmetry group which are relevant
for strongly coupled condensed matter systems, namely, the fermions at
unitarity have been found and they were shown to be easily embedded in string
theory \cite{Son:2008ye,Balasubramanian:2008dm}. Gravitational theories 
having Lifshitz symmetry group relevant for
certain strongly coupled condensed matter systems at their quantum critical
point have also been found \cite{Kachru:2008yh,Taylor:2008tg}, however, their 
embeddings in string theory are not
so easy. In recent literature various methods of embedding the Lifshitz
space-time into string or M-theory have been reported 
\cite{Hartnoll:2009ns}.                       

In this paper, we report on how Lifshitz-like space-time can 
be obtained from certain intersecting brane solutions of string/M theory.
To be precise, we start from the known intersecting 1/4 BPS D1-D5 threshold 
bound state solution of type IIB string theory \cite{Tseytlin:1996bh}. 
We apply two successive
T-duality transformations to it -- first along the common D1-D5 direction to
produce D0-D4 bound state and then along one of the D4-brane directions
to produce D1-D3 bound state. Note that here D1-branes are transverse to 
D3-branes and are delocalized. This is a 1/4 BPS, threshold bound state
unlike the more familiar D1-D3 bound state which is a 1/2 BPS and 
non-threshold bound state \cite{Breckenridge:1996tt,Russo:1996if}. 
An S-duality transformation 
on this D1-D3 bound state will produce F-D3 bound state which is again a 
1/4 BPS, threshold
bound state. Next, application of T-duality along D3-brane directions will
produce F-D2, F-D1 and F-D0 bound states while the application of T-duality
along the common transverse directions of F-strings and D3-branes will
produce F-D4 and F-D5 bound states. Thus we obtain all the F-D$p$ (with
$0\leq p \leq 5$) bound state solutions of type II string theories. Since
these F-D$p$ solutions are U-dual to D1-D5 system, they are 1/4 BPS
and threshold bound states and are different from the known 
1/2 BPS, non-threshold F-D$p$ bound states \cite{Lu:1999uca}.

The near horizon limits of these intersecting F-D$p$ solutions yield
Lifshitz-like space-time\footnote{Such a space-time metric has recently
  been obtained as a holographic dual of some condensed matter system in
  \cite{Ogawa:2011bz,Huijse:2011ef}. Some aspects of these class of 
theories have been 
discussed in \cite{Dong:2012se,Narayan:2012hk,Kim:2012nb}.} in a suitable 
coordinate with the 
dynamical critical exponent $z=2(5-p)/(6-p)$ and the hyperscaling violation 
exponent\footnote{This concept was introduced in random-field Ising system in 
\cite{Fisher:1986zz}. However, in the context of gauge/gravity duality the
hyperscaling violation exponent was identified while describing certain
compressible
metallic states with hidden Fermi surface \cite{Ogawa:2011bz,Huijse:2011ef}, 
where the exponent satisfies $\theta=d-1$, with $d$, the spatial dimensions 
of the boundary 
theory. More general gravity solutions with $\theta$ not satisfying the
relation just mentioned have been discussed in 
\cite{Dong:2012se,Narayan:2012hk,Kim:2012nb}.} 
$\theta=p-(p-2)/(4-p)$ for $p\neq 4$. For $p=4$, the 
near horizon limit does not yield Lifshitz-like space-time, but gives an 
AdS$_2$ space upto a conformal factor. Except for
$p=1$, the dilatons in all these solutions are non-constant and as a
consequence they produce holographic RG flows. The Lifshitz-like solutions 
that we just mentioned are valid in certain range of parameters where the 
effective string coupling and the space-time curvature remain small. 
However, in other regions, we have to either uplift the 
solutions to 11 dimensions or M-theory (for type IIA) or go to the S-dual 
frame (for type IIB). The solutions in other regions also have the 
structures of
Lifshitz-like space-time and we discuss them case by case.
For $p=4$, AdS$_2$ structure is valid in one phase and in other phase we have to
uplift the solution to M-theory, where we get an AdS$_3$ structure,
without a conformal factor. Finally, we also discuss a delocalized F-D1 bound
state solution whose near horizon limit leads to a completely scale invariant
Lifshitz type solution under an asymmetric scale transformation.

This paper is organized as follows. In section 2, we show the construction of
1/4 BPS, threshold F-D$p$ bound states starting from the known D1-D5 system
of type IIB string theory. We then take the near horizon limit which gives
Lifshitz-like space-time and discuss
some generalities. In section 3, we discuss the various solutions case by case
and obtain the phase structures. In section 4, we consider the delocalized 
F-D1 solution and discuss its near horizon structure. Our conclusions 
are presented in section 5.         

\section
{F-D$p$ and the Lifshitz-like space-time} 

We will start from the 1/4 BPS, threshold D1-D5 bound state of type IIB string
theory and then indicate how F-D$p$ bound state can be obtained from there. We
will take the near horizon limit on these solutions and show how Lifshitz-like
space-time appears. We will discuss some generalities for these solutions.

The string metric and the other field configurations of D1-D5 solution
take the following form (see, for example, \cite{Tseytlin:1996bh}),
\bea\label{d1d5}
ds^2 &=& H_1^{\half}H_2^{\half} \left[-H_1^{-1} H_2^{-1} dt^2
+ H_2^{-1}\sum_{i=1}^4 (dx^i)^2 + H_1^{-1}H_2^{-1}(dx^5)^2+ dr^2 + 
r^2 d\Omega_3^2\right]\nn
e^{2\phi} &=& \frac{H_1}{H_2}\nn
A_{[2]} &=& \left(1-H_1^{-1}\right)dt \wedge dx^5, \qquad A_{[6]}\,\,=\,\,
\left(1-H_2^{-1}\right)dt\wedge dx^1 \wedge \cdots \wedge dx^5   
\eea
In the above $H_{1,2}$ are the two harmonic functions given as
\be\label{harmonicd1d5}
H_{1,2} = 1 + \frac{Q_{1,2}}{r^2}
\ee
where $Q_{1,2}$ are the charges associated with D1-branes and D5-branes. 
The radial coordinate transverse to D1-D5 system is given as
$r=\sqrt{(x^6)^2 + \cdots + (x^9)^2}$. We note that D1-branes lie along $x^5$,
whereas D5-branes lie along $x^1,\,x^2,\,\ldots,\,x^5$. The dilaton in
general is not constant and we have put the string coupling $g_s=1$. $A_{[2]}$
and $A_{[6]}$ are the RR 2-form and 6-form which couple to D1-brane and
D5-brane respectively. The constant terms in the form fields are added to
ensure that the solution is asymptotically flat. But when we take the near
horizon limit we generally deal with asymptotically non-flat solutions and in
those cases we will ignore the constant terms in the form fields.

We then apply two successive T-duality, first, along $x^5$ and second, 
along $x^4$ to the above solution and we will get a 1/4 BPS, threshold 
D1-D3 bound state solution of type IIB string theory. The solution
has the form,
\bea\label{d1d3}
ds^2 &=& H_1^{\half}H_2^{\half} \left[-H_1^{-1} H_2^{-1} dt^2
+ H_2^{-1}\sum_{i=1}^3 (dx^i)^2 + H_1^{-1}(dx^4)^2+ dr^2 +
r^2 d\Omega_4^2\right]\nn
e^{2\phi} &=& H_1\nn
A_{[2]} &=& \left(1-H_1^{-1}\right)dt \wedge dx^4, \qquad A_{[4]}\,\,=\,\,
\left(1-H_2^{-1}\right)dt\wedge dx^1 \wedge dx^2 \wedge dx^3
\eea
Note that here D3-branes lie along $x^1,\,x^2,\,x^3$ and are delocalized 
in $x^4$, whereas D1-branes lie
along $x^4$ and are delocalized in $x^1,\,x^2,\,x^3$ directions. Also,
the transverse radial coordinate is given as $r = \sqrt{(x^5)^2 + \cdots
+ (x^9)^2}$ and therefore the harmonic functions have the forms    
\be\label{harmonicd1d3}
H_{1,2} = 1 + \frac{Q_{1,2}}{r^3}  
\ee
$A_{[2]}$ and $A_{[4]}$ are the RR 2-form and 4-form which couple to 
D1-brane and D3-brane respectively and $Q_{1,2}$ are the charges 
associated with them. Although in \eqref{d1d3} the 4-form field has
only electrical component, but it should also include the magnetic
component such that the corresponding field-strength is self-dual.
But we do not write here its exact form. We would also
like to remark that the known D1-D3 bound state of type IIB string 
theory is 1/2 BPS and non-threshold. The solution of the latter type
\cite{Breckenridge:1996tt,Russo:1996if}
also contains a non-zero NSNS $B$-field, which is absent in the above 
solution.

Now an S-duality transformation on this D1-D3 bound state will give
an F-D3 bound state, where `F' denotes the fundamental string and
has the form,
\bea\label{fd3}
ds^2 &=& H_2^{\half} \left[-H_1^{-1} H_2^{-1} dt^2
+ H_2^{-1}\sum_{i=1}^3 (dx^i)^2 + H_1^{-1}(dx^4)^2+ dr^2 +
r^2 d\Omega_4^2\right]\nn
e^{2\phi} &=& H_1^{-1}\nn
B_{[2]} &=& \left(1-H_1^{-1}\right)dt \wedge dx^4, \qquad A_{[4]}\,\,=\,\,
\left(1-H_2^{-1}\right)dt\wedge dx^1 \wedge dx^2 \wedge dx^3
\eea
In \eqref{fd3} $H_{1,2}$ has the same form as given in \eqref{harmonicd1d3},
with $Q_{1,2}$ referring to the charges of F-strings and D3-branes.
D3-branes are along $x^1,\,x^2,\,x^3$ and delocalized in $x^4$.
F-strings are along $x^4$
and delocalized in the D3-brane directions. $B_{[2]}$ is the
NSNS 2-form which couples to F-string and $A_{[4]}$ is the RR 4-form
which couples to D3-brane. We remark that this F-D3 bound state
is 1/4 BPS and threshold unlike the known F-D3 bound state which
is 1/2 BPS and non-threshold \cite{Lu:1999uca}.

Applying a series of T-duality transformations on \eqref{fd3}
along D3-brane directions we get F-D2, F-D1 and F-D0\footnote{There is no 
1/2 BPS F-D0 bound state as is well-known \cite{Lu:1999uca}. In that 
sense this is quite unique in this case. Existence of such state has
been predicted in \cite{Tseytlin:1996hi}.}
and along common transverse directions of F-D3 we get F-D4 and
F-D5\footnote{As we are interested in asymptotically flat solutions we
do not consider F-D6 bound state. Beyond that there are no bound states
in the massless theories.} bound states. All the F-D$p$, with 
$0\leq p \leq 5$, bound state solutions can be written as,
\bea\label{fdp}
ds^2 &=& H_2^{\half} \left[-H_1^{-1} H_2^{-1} dt^2
+ H_2^{-1}\sum_{i=1}^p (dx^i)^2 + H_1^{-1}(dx^{p+1})^2+ dr^2 +
r^2 d\Omega_{7-p}^2\right]\nn
e^{2\phi} &=& \frac{H_2^{\frac{3-p}{2}}}{H_1}\nn
B_{[2]} &=& \left(1-H_1^{-1}\right)dt \wedge dx^{p+1}, \qquad A_{[p+1]}\,\,
=\,\, \left(1-H_2^{-1}\right)dt\wedge dx^1 \wedge \cdots \wedge dx^p
\eea
where the harmonic functions are,
\be\label{harmonicfdp}
H_{1,2} = 1 + \frac{Q_{1,2}}{r^{6-p}}
\ee
with $Q_{1,2}$ representing the charges of F-strings and D$p$-branes. From
\eqref{fdp} it is clear that D$p$-branes lie along $x^1,\,x^2,\cdots,\,x^p$
and are delocalized in the F-string direction, whereas F-strings lie along 
$x^{p+1}$ and are delocalized in the D$p$-brane directions. These are 1/4 BPS, 
threshold bound states and are different from
the known 1/2 BPS, non-threshold F-D$p$ bound states. $B_{[2]}$ is NSNS 2-form
and $A_{[p+1]}$ is the $(p+1)$-form field which couple to F-string and
D$p$-brane respectively.

The near horizon limit of the above F-D$p$ solutions amounts to taking $r\to
0$ limit, such that the harmonic functions in \eqref{harmonicfdp} can be
approximated as
\be\label{approxhar}
H_{1,2} \approx \frac{Q_{1,2}}{r^{6-p}}
\ee
The radial paramater $r$ is holographically related to the RG flow parameter
in the boundary theory and $r \to 0$ corresponds to going to the IR and $r \to
\infty$ corresponds to going to the UV. We will further make a coordinate
transformation $ r \to 1/r$ for convenience and in terms of this new parameter
$ r \to \infty$ ($r \to 0$) corresponds to going to the IR (UV). In terms of
this new $r$ coordinate the metric in \eqref{fdp} reduces in the near horizon
limit to,
\be\label{fdpnh}
ds^2 = Q_2^{\half} r^{\frac{2-p}{2}}\left[-\frac{dt^2}{Q_1Q_2r^{10-2p}}
+ \frac{\sum_i^p(dx^i)^2}{Q_2r^{4-p}} + \frac{(dx^{p+1})^2}{Q_1 r^{4-p}}
+ \frac{dr^2}{r^2} + d\Omega_{7-p}^2\right]
\ee
Further introducing a new coordinate $u$ by the relation
\be\label{newcoord}
u^2 = r^{4-p}
\ee
we can rewtite the metric in \eqref{fdpnh} and other field configurations of
F-D$p$ solutions from \eqref{fdp} as follows,
\bea\label{fdpinnewcoord}
ds^2 &=& Q_2^{\half} u^{\frac{2-p}{4-p}}\left[-\frac{dt^2}{Q_1Q_2
u^{\frac{4(5-p)}{4-p}}}
+ \frac{\sum_i^p(dx^i)^2}{Q_2 u^2} + \frac{(dx^{p+1})^2}{Q_1 u^2}
+ \frac{4}{(4-p)^2}\frac{du^2}{u^2} + d\Omega_{7-p}^2\right]\nn
e^{2\phi} &=& \frac{Q_2^{\frac{3-p}{2}}}{Q_1} u^{\frac{(6-p)(1-p)}{(4-p)}}\nn
B_{[2]} &=& -\frac{1}{Q_1 u^{\frac{2(6-p)}{4-p}}} dt \wedge dx^{p+1},
\qquad A_{[p+1]}\,\, =\,\, -\frac{1}{Q_2 u^{\frac{2(6-p)}{4-p}}} dt \wedge dx^1
\wedge \cdots \wedge dx^p
\eea 
Note that the coordinate relation defined in \eqref{newcoord} works for all
$p$ except for $p=4$ and so, the field configuration given in
\eqref{fdpinnewcoord} is valid for all $p \neq 4$. Therefore, $p=4$ case needs
to be discussed separately. This we will do in the next section where
various RG flow and the phase structure will be considered case by case. 
Because of the
relation \eqref{newcoord} it is clear that for $p<4$, $r \to \infty$ implies 
$u \to
\infty$ and corresponds to going to the IR, whereas $r \to 0$ implies $u\to
0$ and this corresponds to going to the UV. On the other hand for $p>4$, 
$r \to \infty$ implies $u \to 0$ and corresponds to going to the IR, whereas, 
$r \to 0$ implies $u \to \infty$ and corresponds to going to the UV.
From \eqref{fdpinnewcoord} we observe that under the following scale
transformations
\be\label{scaling}
t \to \lambda^{\frac{2(5-p)}{4-p}} t \equiv \lambda^z t, 
\quad x^{1,2,\ldots,(p+1)} \to \lambda x^{1,2,\ldots,(p+1)}, 
\quad u \to \lambda u
\ee
only the part in the square bracket of the metric remains invariant. However,
the full metric changes. Now instead of looking at the full metric, if we
compactify the theory on S$^{7-p}$ and write the reduced metric in Einstein
frame it takes the form,
\be\label{emetric}
ds_{p+3}^2 = Q_1^{\frac{2}{p+1}} Q_2 u^{2[\frac{p(4-p)-(p-2)}{(4-p)(p+1)}]}
\left[-\frac{dt^2}{Q_1Q_2
u^{\frac{4(5-p)}{4-p}}}
+ \frac{\sum_i^p(dx^i)^2}{Q_2 u^2} + \frac{(dx^{p+1})^2}{Q_1 u^2}
+ \frac{4}{(4-p)^2}\frac{du^2}{u^2}\right]
\ee
Under the scaling \eqref{scaling} this metric changes as,
\be\label{hscaling}
ds_{p+3} \to \lambda^{\frac{p(4-p)-(p-2)}{(4-p)(p+1)}}ds_{p+3} \equiv 
\lambda^{\theta/d} ds_{p+3}
\ee
where $z$ in \eqref{scaling} is called the dynamical critical exponent and 
$\theta$ in \eqref{hscaling} is called the hyperscaling violation exponent.
$d$ is the spatial dimension of the boundary theory which is $(p+1)$ in this
case. We thus find that the near horizon geometries of the F-D$p$ bound states
produce Lifshitz-like theories with dynamical critical exponent $z$ and
hyperscaling violation exponent $\theta$ having values,
\be\label{schsc}
z =  \frac{2(5-p)}{4-p}, \qquad \theta = p - \frac{p-2}{4-p}
\ee
$z$ takes integer values 3, 4 and 0 for F-D2, F-D3 and F-D5
solutions and $\theta$ takes the value 2 for the first two cases and 8 for
the last case. From \eqref{fdpinnewcoord} we note that
the dilaton is constant only for F-D1 solution and for other solutions it
varies with $u$.

Metric of the type given in \eqref{fdpinnewcoord} (or the compactified
version of it\footnote{Note that it is the compact metric whose scaling
  property defines the hyperscaling violation exponent and not the full
  ten/eleven dimensional metric.}) has recently been found 
\cite{Huijse:2011ef} 
to be useful in describing some condensed matter system. In fact, it has been
observed that some non-Fermi liquid metallic states with hidden Fermi surface
can be described by a holographic IR metric with a dynamical critical exponent
$z$ and a hyperscaling violation exponent $\theta$. However, since a
consistent gravity theory must satisfy the null energy condition (NEC), the 
pairs ($z,\theta$) which satisfy NEC given by \cite{Dong:2012se},
\bea\label{null}
(d-\theta)(d(z-1)-\theta) &\geq& 0\nn
(z-1)(d+z-\theta) &\geq& 0
\eea 
will therefore lead to a consistent dual field theory. It can be easily
checked that the pairs ($z,\theta$) obtained in \eqref{schsc} indeed satisfy
the NEC \eqref{null}. Other string theoretic realization of such 
metric has been
obtained in \cite{Narayan:2012hk}. See also \cite{Singh:2012un} for some
other constructions.  
 
Under the scaling \eqref{scaling} the dilaton and the form 
fields change as,
\be\label{change}
\phi \to \phi + \frac{(6-p)(1-p)}{2(4-p)} \log \lambda, \quad
B_{[2]} \to \lambda^{\frac{2-p}{4-p}} B_{[2]}, \quad A_{[p+1]} \to 
\lambda^{\frac{2-(p-2)^2}{4-p}} A_{[p+1]}
\ee
This shows that $B_{[2]}$ remains invariant under the scaling only for F-D2
solution but $A_{[p+1]}$ is never invariant. In section 4, we
will discuss a case where the full solution will remain invariant under
a scale transformation without any hyperscaling violation. Note that the 
Lifshitz-like solutions \eqref{fdpinnewcoord} we obtained preserve at
least a 1/4 space-time SUSY as the intersecting solutions we started out 
with are 1/4 BPS.

\section{RG flow \& phase structure: case by case 
study} 

Since the RG flows and the phase structures are quite different for different 
values of $p$, we will study them case by case in this section.

\subsection{$p=0$: F-D0 case}

The near horizon limit in this case gives the following field
configurations with a Lifshitz-like space-time (see \eqref{fdpinnewcoord}), 
\bea\label{fd0innewcoord}
ds^2 &=& Q_2^{\half} u^{\half}\left[-\frac{dt^2}{Q_1Q_2
u^5}
+ \frac{(dx^1)^2}{Q_1 u^2} + \frac{1}{4}\frac{du^2}{u^2} + 
d\Omega_{7}^2\right]\nn
e^{2\phi} &=& \frac{Q_2^{\frac{3}{2}}}{Q_1} u^{\frac{3}{2}}\nn
B_{[2]} &=& -\frac{1}{Q_1 u^3} dt \wedge dx^1,
\qquad A_{[1]}\,\, =\,\, -\frac{1}{Q_2 u^3} dt
\eea
We have already discussed the scaling properties of this solution in
section 2. Here we will discuss its RG flow and the phase structure. 
Note that the above gravity description is valid when the effective string
coupling $e^{\phi}$ and the curvature of the space-time remains small. 
From \eqref{fd0innewcoord} we find that they amount to the following range
of $u$ where the above gravity description can be trusted,
\be\label{rangefd0}
\frac{1}{Q_2} \ll u \ll \frac{Q_1^{\frac{2}{3}}}{Q_2}
\ee
However when $u \geq Q_1^{2/3}/Q_2$, the dilaton becomes large and 
the gravity description breaks down and we have to uplift the solution
to eleven dimensions. In eleven dimensions the solution takes the form,
\bea\label{mw}
ds^2 &=& Q_1^{\frac{1}{3}}\left[-\frac{dt^2}{Q_1Q_2
u^5}
+ \frac{(dx^1)^2}{Q_1 u^2} + \frac{Q_2}{Q_1}u\left(dx^{11} - \frac{1}{Q_2u^3}dt
\right)^2 + \frac{1}{4}\frac{du^2}{u^2} + 
d\Omega_{7}^2\right]\nn
A_{[3]} &=& -\frac{1}{Q_1 u^3} dt \wedge dx^1 \wedge dx^{11}
\eea   
In M-theory the above solution represents the near horizon limit of a 1/4 BPS, 
threshold bound state of an M2-brane (along $x^1,\,x^{11}$) with a wave 
along $x^{11}$ \cite{Tseytlin:1996hi}. Note that under the following 
scale transformation 
\be\label{scalingmw}
t \to \lambda^{5/2} t, \quad x^1 \to \lambda x^1, \quad u \to \lambda u,
\quad x^{11} \to \lambda^{-1/2} x^{11}
\ee
both the metric and the form field in \eqref{mw} remain
invariant. Thus here also we get a Lifshitz space-time with dynamical
critical exponent $z=5/2$ and no hyperscaling violation. However, we have an
asymmetric scaling of $x^1$ and $x^{11}$ in this case. The gravity description
remains valid when the eleven dimensional metric has small curvature in Planck
unit, i.e., $Q_1 \gg 1$.

\subsection{$p=1$: F-D1 case}

As we have observed in section 2, the dilaton in this case remains
constant and therefore there is no holographic RG flow. The near horizon
configuration has the form,
\bea\label{fd1innewcoord}
ds^2 &=& Q_2^{\half} u^{\frac{1}{3}}\left[-\frac{dt^2}{Q_1Q_2
u^{\frac{16}{3}}}
+ \frac{(dx^1)^2}{Q_2 u^2} + \frac{(dx^2)^2}{Q_1 u^2} + 
\frac{4}{9}\frac{du^2}{u^2} + 
d\Omega_{6}^2\right]\nn
e^{2\phi} &=& \frac{Q_2}{Q_1}\nn
B_{[2]} &=& -\frac{1}{Q_1 u^{\frac{10}{3}}} dt \wedge dx^2,
\qquad A_{[2]}\,\, =\,\, -\frac{1}{Q_2 u^{\frac{10}{3}}} dt \wedge dx^1
\eea
We have discussed the scaling properties of this solution in section 2.
Here we note that for the above gravity description to remain valid
the effective string coupling $e^{\phi}$ and the curvature must remain small.
In this case those amount to,
\be\label{rangefd1}
u \gg \frac{1}{Q_2^{\frac{3}{2}}} \gg \frac{1}{Q_1^{\frac{3}{2}}} 
\ee 
However, for the case $Q_2/Q_1 \geq 1$, $e^{\phi}$ becomes large and we have 
to go to the S-dual frame. The S-dual frame configuration representing the 
near horizon limit of D1-F bound state will be given as,
\bea\label{fd1sdual}
ds^2 &=& Q_1^{\half} u^{\frac{1}{3}}\left[-\frac{dt^2}{Q_1Q_2
u^{\frac{16}{3}}}
+ \frac{(dx^1)^2}{Q_2 u^2} + \frac{(dx^2)^2}{Q_1 u^2} + 
\frac{4}{9}\frac{du^2}{u^2} + 
d\Omega_{6}^2\right]\nn
e^{2\phi} &=& \frac{Q_1}{Q_2}\nn
B_{[2]} &=& -\frac{1}{Q_2 u^{\frac{10}{3}}} dt \wedge dx^1,
\qquad A_{[2]}\,\, =\,\, -\frac{1}{Q_1 u^{\frac{10}{3}}} dt \wedge dx^2
\eea 
We again get Lifshitz-like space-time with the same scaling property as the
original solution \eqref{fd1innewcoord}. In order to trust the S-dual gravity
configuration we must have,
\be\label{ranged1f}
u \gg \frac{1}{Q_1^{\frac{3}{2}}} \gg \frac{1}{Q_2^{\frac{3}{2}}} 
\ee

\subsection{$p=2$: F-D2 case}

The field configurations in the near horizon limit in this case have 
the form,
\bea\label{fd2innewcoord}
ds^2 &=& Q_2^{\half} \left[-\frac{dt^2}{Q_1Q_2
u^6}
+ \frac{\sum_{i=1}^2 (dx^i)^2}{Q_2 u^2} + \frac{(dx^3)^2}{Q_1 u^2} + 
\frac{4}{9}\frac{du^2}{u^2} + 
d\Omega_{5}^2\right]\nn
e^{2\phi} &=& \frac{Q_2^{\half}}{Q_1 u^2}\nn
B_{[2]} &=& -\frac{1}{Q_1 u^4} dt \wedge dx^3,
\qquad A_{[3]}\,\, =\,\, -\frac{1}{Q_2 u^{4}} dt \wedge dx^1 \wedge dx^2
\eea
The metric has a Lifshitz-like structure and the scaling property of this 
solution is described earlier. We remark that unlike in other
F-D$p$ cases, here the full metric remains invariant under the scale
transformations and so, one might think that this case gives Lifshitz
space-time (without any hyperscaling violation), but this is not true.
The reason is that the dilaton is not constant. Therefore, when one
compactifies the theory on S$^5$, and writes the 5-dimensional metric in
the Einstein frame, the resulting metric will not remain invariant under the
scaling and will give rise to a hyperscaling violation. Now
for the gravity description \eqref{fd2innewcoord} to remain valid we must 
impose the conditions that
the effective string coupling $e^{\phi}$ and the curvature remain small.
In this case they amount to the following condition on $u$,
\be\label{rangefd2}
u \gg \frac{Q_2^{\frac{1}{4}}}{Q_1^{\half}}, \qquad {\rm along\,\, with} \qquad
Q_2 \gg 1
\ee 
However when $u \leq Q_2^{1/4}/Q_1^{1/2}$, the effective string coupling
$e^{\phi}$ becomes large and we have to uplift the
solution to M-theory. The uplifted solution has the form,
\bea\label{fd2mtheory}
ds^2 &=& Q_1^{\frac{1}{3}}Q_2^{\frac{1}{3}} u^{\frac{2}{3}} 
\left[-\frac{dt^2}{Q_1Q_2
u^6}
+ \frac{\sum_{i=1}^2 (dx^i)^2}{Q_2 u^2} + \frac{(dx^3)^2 + 
(dx^{11})^2}{Q_1 u^2} + \frac{du^2}{u^2} + 
d\Omega_{5}^2\right]\nn
A_{[3]} &=& -\frac{1}{Q_1 u^4} dt \wedge dx^3 \wedge dx^{11},
\qquad A'_{[3]}\,\, =\,\, -\frac{1}{Q_2 u^{4}} dt \wedge dx^1 \wedge dx^2
\eea
The above configuration represents the near horizon limit of two intersecting
M2-branes \cite{Tseytlin:1996bh} one along $x^1,\,x^2$ and the 
other along $x^3,\,x^{11}$. Under the scaling
\be\label{fd2mscaling} 
t \to \lambda^3 t, \quad x^{1,2,3,11} \to \lambda x^{1,2,3,11}, \quad u \to
\lambda u
\ee 
the part of the metric in the square bracket remains invariant. However, the
metric compactified on S$^5$ in Einstein frame and the other fields 
transform as,
\be\label{fd2mtheoryscaling}
ds_6 \to \lambda^{\frac{3}{4}}ds_6, \quad A_{[3]} \to \lambda A_{[3]}, \quad
A'_{[3]} \to \lambda A'_{[3]}
\ee
We thus find that this theory also has a Lifshitz-like structure with the
dynamical scaling exponent $z=3$ and the hyperscaling violation exponent
$\theta = 3$, where this pair of $(z,\theta)$ satisfies the NEC \eqref{null}.
The gravity description \eqref{fd2mtheory} can be trusted for $u \gg
1/\sqrt{Q_1Q_2}$. 

\subsection{$p=3$: F-D3 case}

In this case the metric having a Lifshitz-like structure and the other field
configurations in the near horizon limit are given as,
\bea\label{fd3innewcoord}
ds^2 &=& Q_2^{\half} \frac{1}{u}\left[-\frac{dt^2}{Q_1Q_2
u^8}
+ \frac{\sum_{i=1}^3 (dx^i)^2}{Q_2 u^2} + \frac{(dx^4)^2}{Q_1 u^2} + 
4\frac{du^2}{u^2} + 
d\Omega_{4}^2\right]\nn
e^{2\phi} &=& \frac{1}{Q_1 u^6}\nn
B_{[2]} &=& -\frac{1}{Q_1 u^6} dt \wedge dx^4,
\qquad F_{[5]}\,\, =\,\, (1+\ast) \frac{6}{Q_2 u^7} du \wedge dt \wedge dx^1 
\wedge dx^2 \wedge dx^3
\eea 
In \eqref{fd3innewcoord} instead of the 4-form gauge field we have given 
the self-dual 5-form field strength which couples to D3-brane. The scaling 
property of this solution has already been discussed in section 2.
Here we note that the gravity description remains valid only when the
effective string coupling $e^{\phi}$ and the curvature of space-time remain
small. In this case we get the following range of $u$ where both the
conditions are satisfied,
\be\label{rangefd3}
\frac{1}{Q_1^{\frac{1}{6}}} \ll u \ll Q_2^{\half}
\ee
For $u \leq 1/Q_1^{1/6}$ we have to go to the S-dual frame. The S-dual
configurations representing the near horizon limit of D1-D3 bound state have
the form,
\bea\label{fd3sdual}
ds^2 &=& Q_1^{\half} Q_2^{\half} u^2\left[-\frac{dt^2}{Q_1Q_2
u^8}
+ \frac{\sum_{i=1}^3 (dx^i)^2}{Q_2 u^2} + \frac{(dx^4)^2}{Q_1 u^2} + 
4\frac{du^2}{u^2} + 
d\Omega_{4}^2\right]\nn
e^{2\phi} &=& Q_1 u^6\nn
A_{[2]} &=& -\frac{1}{Q_1 u^6} dt \wedge dx^4,
\qquad F_{[5]}\,\, =\,\, (1+\ast) \frac{6}{Q_2 u^7} du \wedge dt \wedge dx^1 
\wedge dx^2 \wedge dx^3
\eea 
Under the scaling
\be\label{d1d3scaling} 
t \to \lambda^4 t, \quad x^{1,2,3,4} \to \lambda x^{1,2,3,4}, \quad
u \to \lambda u
\ee
the Einstein frame metric after an S$^4$ compactification and the various 
fields transform as,
\be\label{d1d3change}
ds_6 \to \lambda^{\half} ds_6, \quad \phi \to \phi + 3 \log \lambda, 
\quad A_{[2]} \to 
\lambda^{-1} A_{[2]}, \quad F_{[5]} \to \lambda F_{[5]}
\ee
We thus find that the S-dual configurations also have 
a Lifshitz-like space-time
with the same dynamical scaling exponent $z=4$ and the hyperscaling violation
exponent $\theta = 2$ as the original theory and therefore satisfies the 
NEC \eqref{null}. The S-dual
gravity description can be trusted in the range $1/(Q_1Q_2)^{1/4} \ll u \ll
1/Q_1^{1/6}$.  

\subsection{$p=5$: F-D5 case}

The near horizon limit of F-D5 bound state has a Lifshitz-like space-time
along with other field configurations given by,
\bea\label{fd5innewcoord}
ds^2 &=& Q_2^{\half} u^3\left[-\frac{dt^2}{Q_1Q_2}
+ \frac{\sum_{i=1}^5 (dx^i)^2}{Q_2 u^2} + \frac{(dx^6)^2}{Q_1 u^2} + 
4\frac{du^2}{u^2} + 
d\Omega_{2}^2\right]\nn
e^{2\phi} &=& \frac{u^4}{Q_1 Q_2}\nn
B_{[2]} &=& -\frac{u^2}{Q_1} dt \wedge dx^6,
\qquad A_{[6]}\,\, =\,\, -\frac{u^2}{Q_2} dt \wedge dx^1 
\wedge \cdots \wedge dx^5
\eea 
We have observed the scaling properties of the various fields in section 2.
Here we will study its RG flow and the phase structure. From 
\eqref{fd5innewcoord} we notice
that the above gravity description is valid when $e^{\phi}$ and the curvature
remain small which amounts to the following range of $u$,
\be\label{rangefd5}
Q_2^{\frac{1}{6}} \ll u \ll (Q_1Q_2)^{\frac{1}{4}}
\ee
However, for $u \geq (Q_1Q_2)^{1/4}$,
effective string coupling $e^{\phi}$ becomes large and the gravity description 
breaks down. In
that case we have to go to the S-dual frame. The S-dual configuration in this
case would be given by the near horizon limit of D1-NS5 bound state and has
the form,
\bea\label{fd5sdual}
ds^2 &=& Q_1^{\half} Q_2 u\left[-\frac{dt^2}{Q_1Q_2}
+ \frac{\sum_{i=1}^5 (dx^i)^2}{Q_2 u^2} + \frac{(dx^6)^2}{Q_1 u^2} + 
4\frac{du^2}{u^2} + d\Omega_{2}^2\right]\nn
e^{2\phi} &=& \frac{Q_1 Q_2}{u^4}\nn
A_{[2]} &=& -\frac{u^2}{Q_1} dt \wedge dx^6,
\qquad H_{[3]}\,\, =\,\, -Q_2 dx^6 \wedge \epsilon_2 
\eea
where $\epsilon_2$ is the volume form of a unit two-sphere. From 
\eqref{fd5sdual} we find that under the scaling 
\be\label{scalingfd5}
t \to \lambda^0 t, \quad x^{1,2,\ldots,6} \to \lambda x^{1,2,\ldots,6},
\quad u \to \lambda u
\ee
the Einstein frame metric on S$^2$ compactification and the
various other fields transform as,
\be\label{changefd5}
ds_8 \to \lambda^{\frac{4}{3}} ds_8, \quad \phi \to \phi - 2\log\lambda, \quad
A_{[2]} \to \lambda^3 A_{[2]}, \quad H_{[3]} \to \lambda H_{[3]}
\ee
We therefore find that the theory in the UV also has a Lifshitz-like space-time
with the same dynamical scaling exponent $z=0$ and the hyperscaling violation 
exponent $\theta = 8$ as the original theory. This pair of $(z,\theta)$ 
satisfies the NEC \eqref{null} as we have noted before. The gravity 
description in this case can be trusted for $u \gg
(Q_1Q_2)^{1/4}$, where the effective string coupling and the curvature remain 
small.

\subsection{$p=4$: F-D4 case}

We have mentioned before that $p=4$ case is special since in this case
the introduction of new coordinate $u$ is not possible (see \eqref{newcoord}).
So, we have to write the near horizon configuration of F-D4 bound state in
terms of the original coordinate $r$. From the general F-D$p$ solution
\eqref{fdp} and using the near horizon limit of the harmonic functions
\eqref{approxhar} and further making the transformation $r \to 1/r$, we 
can write the F-D4 solution in the near horizon limit as,
\bea\label{fd4}
ds^2 &=& \frac{Q_2^{\half}}{r} \left[-\frac{dt^2}{Q_1Q_2 r^2}
+ \frac{\sum_{i=1}^4 (dx^i)^2}{Q_2} + \frac{(dx^5)^2}{Q_1} + 
\frac{dr^2}{r^2} + 
d\Omega_{3}^2\right]\nn
e^{2\phi} &=& \frac{1}{Q_1 Q_2^{\half} r^3}\nn
B_{[2]} &=& -\frac{1}{Q_1 r^2} dt \wedge dx^5,
\qquad A_{[5]}\,\, =\,\, -\frac{1}{Q_2 r^2} dt \wedge dx^1 
\wedge \cdots \wedge dx^4
\eea
So, it is clear that we do not get a Lifshitz-like space-time from F-D4,
however, the near horizon metric has the structure of AdS$_2$ space upto a
conformal factor. The dilaton is non-constant and therefore will produce
a holographic RG flow. The above gravity description is valid when
the effective string coupling $e^{\phi}$ and the curvature remain
small. In this case these amount to the following range of $r$,
\be\label{rangefd4}
\frac{1}{Q_1^{\frac{1}{3}} Q_2^{\frac{1}{6}}} \ll r \ll Q_2^{\half}
\ee
However when $r \leq 1/(Q_1^{1/3} Q_2^{1/6})$, we have to uplift the 
solution to M-theory.
The eleven dimensional metric has the form,
\be\label{fd4mtheory}
ds^2 = Q_1^{\frac{1}{3}} Q_2^{\frac{2}{3}}\left[-\frac{dt^2}{Q_1 Q_2 r^2}
+ \frac{\sum_{i=1}^4 (dx^i)^2}{Q_2} + \frac{(dx^5)^2}{Q_1} +
\frac{(dx^{11})^2}{Q_1Q_2r^2} + \frac{dr^2}{r^2} + d\Omega_3^2\right]
\ee
The above solution represents the near horizon limit of intersecting M2-M5
brane meeting on a string where M2-branes are along $x^5$ and $x^{11}$ and
M5-branes are along $x^1,\,\ldots,\,x^4,\,x^{11}$. Note that this uplifted
solution has the structure of AdS$_3$ space without any conformal 
factor\footnote{This solution has previously been obtained in 
\cite{Kallosh:1997qw}.}.
Thus the UV theory has an AdS$_3$ structure. This gravity description can be
trusted as long as $Q_1^{1/6} Q_2^{1/3} \gg 1$.

\section{A delocalized F-D1 and Lifshitz space-time} 
  
In section 3, we noted that among all the F-D$p$ solutions only
F-D2 leads to fully scale invariant ten-dimensional metric in the near horizon 
limit without any 
conformal factor. However, the dilaton as well as the RR form field do 
not remain invariant under the scale transformation. Only the NSNS form 
field 
remains invariant. F-D1, on the other hand, leads to constant dilaton in the
near horizon limit and therefore remains invariant under the scale
transformation.
However, the full metric, the NSNS as well as the RR form fields do not 
remain scale invariant. In this
section we will describe a solution which is somewhat in between the two
solutions we decribed, namely, a delocalized F-D1 solution whose near horizon
limit will lead to (asymmetric) Lifshitz space-time and the other fields 
will be invariant under the scale transformation.  

To obtain the delocalized solution we start from the F-D2 solution
given in \eqref{fdp} for $p=2$ and then apply T-duality along one of the brane
directions of the D2-brane. This will produce a delocalized\footnote{ 
Note that a localized solution is usually obtained from a delocalized one
when we replace the extended source of the delocalized solution by a point 
source and this is the usual procedure when one
takes T-duality. However, instead of changing the source if we keep the
extended source the solution remains delocalized 
\cite{Lu:1995cs,Breckenridge:1996tt}.} 
F-D1 solution. To get its
form let us first write F-D2 solution from \eqref{fdp} as,
\bea\label{fd2}
ds^2 &=& H_2^{\half} \left[-H_1^{-1} H_2^{-1} dt^2
+ H_2^{-1}\sum_{i=1}^2 (dx^i)^2 + H_1^{-1}(dx^{3})^2+ dr^2 +
r^2 d\Omega_{5}^2\right]\nn
e^{2\phi} &=& \frac{H_2^{\frac{1}{2}}}{H_1}\nn
B_{[2]} &=& \left(1-H_1^{-1}\right)dt \wedge dx^{3}, \qquad A_{[3]}\,\,
=\,\, \left(1-H_2^{-1}\right)dt\wedge dx^1 \wedge dx^2
\eea 
where the harmonic functions are given as
\be\label{harmonicfd2}
H_{1,2} = 1 + \frac{Q_{1,2}}{r^4}
\ee
Taking T-duality along $x^2$ and then renaming $x^2 \leftrightarrow x^3$ we
get the delocalized F-D1 solution as,
\bea\label{fd1delocalized}
ds^2 &=& H_2^{\half} \left[-H_1^{-1} H_2^{-1} dt^2
+ H_2^{-1} (dx^1)^2 + H_1^{-1}(dx^{2})^2+  (dx^3)^2+ dr^2 +
r^2 d\Omega_{5}^2\right]\nn
e^{2\phi} &=& \frac{H_2}{H_1}\nn
B_{[2]} &=& \left(1-H_1^{-1}\right)dt \wedge dx^{2}, \qquad A_{[2]}\,\,
=\,\, \left(1-H_2^{-1}\right)dt\wedge dx^1
\eea 
with the harmonic functions having the same form as given in 
\eqref{harmonicfd2}. It is clear that the F-strings lie along $x^2$ and
are delocalized along $x^1,\,x^3$, whereas, D1-branes lie along $x^1$ and
are delocalized along $x^2,\,x^3$. So, $x^3$ is the common delocalized
direction. Now if we go to the near horizon limit ($r \to 0$) by approximating
the harmonic functions by $H_{1,2} \approx Q_{1,2}/r^4$ and then make a change
of coordinates $r \to 1/r$, the above solution \eqref{fd1delocalized}
reduces to 
\bea\label{fd1deloc}
ds^2 &=& Q_2^{\half} \left[-\frac{dt^2}{Q_1Q_2 r^6}
+ \frac{(dx^1)^2}{Q_2 r^2} + \frac{(dx^2)^2}{Q_1 r^2} + r^2 (dx^3)^2 + 
\frac{dr^2}{r^2} + d\Omega_{3}^2\right]\nn
e^{2\phi} &=& \frac{Q_2}{Q_1}\nn
B_{[2]} &=& -\frac{1}{Q_1 r^4} dt \wedge dx^2,
\qquad A_{[2]}\,\, =\,\, -\frac{1}{Q_2 r^4} dt \wedge dx^1 
\eea
The solution \eqref{fd1deloc} is invariant under the following scaling,
\be\label{fd1delocscaling}
t \to \lambda^3 t, \quad x^{1,\,2} \to \lambda x^{1,\,2}, \quad x^3 \to
\lambda^{-1} x^3, \quad r \to \lambda r
\ee
We thus find an asymmetric (since $x^3$ transforms differently than
$x^{1,2}$) Lifshitz space-time with the dynamical scaling exponent $z=3$ and
no hyperscaling violation. Note that the dilaton and the other form fields
also remain invariant under the scale transformation \eqref{fd1delocscaling}.

The above gravity solution is valid when $e^{\phi} = \sqrt{Q_2/Q_1} \ll 1$
and also $Q_2^{1/2} \gg 1$.
However if $Q_2/Q_1 \gg 1$, we have to go to the S-dual frame. The S-dual
solution also has a very similar form, as we have discussed in section 3, with 
the same scaling property as the original solution. 

\section{Conclusion}

To summarize, in this paper we have shown how to construct 1/4 BPS, threshold
F-D$p$ (with $0\leq p \leq 5$) bound states of type II string theories 
starting from the well-known 
D1-D5 system of type IIB string theory by applying two T-duality, an 
S-duality and then a series of T-duality transformations. The near horizon
limits of these solutions (for $p\neq 4$) give rise to Lifshitz-like 
space-time with the
dynamical critical exponent $z=2(5-p)/(4-p)$ and the hyperscaling violation
exponent $\theta = p - (p-2)/(4-p)$ in a suitable coordinate. We have
checked that these values of $(z,\theta)$ satisfy the null energy
condition given in \eqref{null}. As a consistent gravity theory must 
satisfy the null energy 
condition in terms of $(z,\theta)$, the pairs which
satisfy these conditions will therefore lead to a physically sensible dual
field theory. The dilatons are in general 
non-constant except for $p=1$ and will generate holographic RG flow. 
We have given the scaling properties of the
dilatons as well as the other form-fields. We have discussed the phase
structures of various theories case by case. At different regions of the
RG flow parameter, there are different phases. We have analyzed the scaling
properties of the theories in other phases and found that they also have
Lifshitz-like structure with different dynamical critical exponents and
hyperscaling violation exponents. In all cases they also satisfy the null
energy condition \eqref{null} as discussed in section 3.    
For $p=4$, we did not get Lifshitz-like space-time, but the near horizon
geometry in this case has an AdS$_2$ structure upto a conformal factor. 
In the strongly coupled phase the geometry has an AdS$_3$ structure without 
the conformal factor. 
We have also discussed a case of a delocalized F-D1 bound state. Here the 
whole near horizon solution is invariant under an asymmetric scaling of the
coordinates. All these solutions discussed here are supersymmetric as they are
obtained from 1/4 BPS string states.

The gravity solutions with Lifshitz scaling along with hyperscaling violation, 
i.e., the type we have
discussed in this paper (including the near horizon limit of (F, D2) solution
and its M-theory lift which has $\theta = d-1$ (see subsection 3.3)) have been
used before to model certain strongly interacting condensed matter system with
Fermi surface. As at weak coupling it is known that a Fermi surface can be 
obtained by deforming a relativistic theory with a non-zero chemical 
potential \cite{Huijse:2011hp}, it would be interesting to see whether 
the gravity solutions we have
obtained in this paper can also be obtained as some kind of deformation of 
certain relativistic solutions. Also note that among the various scaling
symmetries obtained in sections 2 -- 4, the ones discussed in subsection 3.5
(for (F, D5) and its S-dual case in \eqref{scalingfd5}), subsection 3.1 (for
M-theory lift of (F, D0) in \eqref{scalingmw}) and section 4 (for delocalized
(F, D1) in \eqref{fd1delocscaling}) are quite unusual. In \eqref{scalingfd5} we
found $z=0$ which appears to imply that there is no relaxation in time for 
the system 
described by the boundary theory. On the other hand for \eqref{scalingmw} and
\eqref{fd1delocscaling} we found negative scaling exponents for some boundary
coordinates and this apparently would imply critical speeding up of the 
system in those directions. It would be interesting to 
understand the field theoretic meaning of these scaling symmetries along with
the other solutions.

\vspace{.2cm}
\section*{note added:}

After we submitted the paper to the archive we received some comments from
K. Narayan which has helped us to properly identify the hyperscaling
violation exponent. We are grateful to him for pointing this out to us.

\vspace{.2cm}

\section*{Acknowledgements}
One of the authors (PD) would like to acknowledge thankfully the financial
support of the Council of Scientific and Industrial Research, India
(SPM-07/489 (0089)/2010-EMR-I).
The other author (SR) would like to thank Shankhadeep Chakrabortty for 
asking some questions which prompted us to construct the F-D$p$ solutions 
discussed here. We would also like to thank Harvendra Singh for discussions.
We would like to thank the anonymous referee for comments which has helped,
we hope, to improve the manuscript.

\end{document}